\documentclass[preprint, double-spaced,showkeys,showpacs]{revtex4}
\usepackage{graphicx,amsmath,bbm,bm,array,subfigure} 
\usepackage{appendix}
\usepackage{lipsum}
\usepackage[linktocpage=true,colorlinks=true,linkcolor=blue,citecolor=blue,urlcolor=blue]{hyperref}
\usepackage{MnSymbol}
\def\nn{\nonumber\\}

\begin{document}
\title{Transmission of airborne virus through sneezed and coughed droplets}
\author{Santosh K. Das}
\email{santosh@iitgoa.ac.in}
\affiliation{School of Physical Sciences, Indian Institute of Technology Goa, Ponda-403401, Goa, India}
\author{Jan-e Alam}
\email{jane@vecc.gov.in}
\affiliation{Variable Energy Cyclotron Centre, 1/AF Bidhan Nagar, Kolkata- 700064, India}
\affiliation{Homi Bhabha National Institute, Training School Complex, Mumbai - 400085, India}
\author{Salvatore Plumari}
\email{salvatore.plumari@hotmail.it }
\affiliation{Department of Physics and Astronomy, University of Catania, 
Via S. Sofia 64, I-95125 Catania, Italy}
\affiliation{Laboratori Nazionali del Sud, INFN-LNS, Via S. Sofia 62, I-95123 Catania, Italy}
\author{Vincenzo Greco}
\email{greco@lns.infn.it }
\affiliation{Department of Physics and Astronomy, University of Catania, 
Via S. Sofia 64, I-95125 Catania, Italy}
\affiliation{Laboratori Nazionali del Sud, INFN-LNS, Via S. Sofia 62, I-95123 Catania, Italy}
\def\zbf#1{{\bf {#1}}}
\def\bfm#1{\mbox{\boldmath $#1$}}
\def\hf{\frac{1}{2}}
\def\sl{\hspace{-0.15cm}/}
\def\omit#1{_{\!\rlap{$\scriptscriptstyle \backslash$}
{\scriptscriptstyle #1}}}
\def\vec#1{\mathchoice
        {\mbox{\boldmath $#1$}}
        {\mbox{\boldmath $#1$}}
        {\mbox{\boldmath $\scriptstyle #1$}}
        {\mbox{\boldmath $\scriptscriptstyle #1$}}
}
\def \beq{\begin{equation}}
\def \eeq{\end{equation}}
\def \beqa{\begin{eqnarray}}
\def \eeqa{\end{eqnarray}}
\def \nn{\nonumber}
\def \pd{\partial}

\def \la{\langle}
\def \ra{\rangle}

\begin{abstract}
The spread of COVID19 through  droplets ejected by infected individuals
during sneezing and coughing has been considered as a matter of key concern.
Therefore, a quantitative understanding of the propagation of droplets containing virus
assumes immense importance.  Here we  investigate the evolution of droplets in space and time
under varying external conditions of temperature, humidity and wind flow  by using
laws of statistical and fluid mechanics.  The effects of
drag, diffusion and the gravity on  droplets of
different sizes and ejection velocities have been considered
during their motion in the air. In still air we found that bigger droplets
traverse larger distance but the smaller droplets remain suspended in the air for longer time.
So, in still air the horizontal distance that a healthy individual should maintain from
an infected one is determined by the bigger droplets but the time interval to be maintained
is determined by the smaller droplets.
We show that in places with wind flow the lighter droplets
travel larger distance and  remain suspended in the air for longer time.
Therefore, we conclude that both temporal and the geometric distance that
a healthy individual should maintain from an infected one is
determined by the smaller droplets under flowing air which makes 
the use of mask  mandatory to prevent the virus.
The maintenance of only stationary separation between healthy
and infected individuals is not substantiated.
The quantitative results  obtained here will be useful to devise strategies for
preventing the spread of other types of droplets also containing microorganisms.

\end{abstract}

\maketitle
\section{Introduction}
It is common knowledge that droplets released through coughing, sneezing,
speaking or breathing contain microorganism (bacteria, virus, fungi, etc) causing
a large number of diseases~\cite{Wells}. 
The droplets containing pathogens~\cite{cflugge} can transmit from
an infected individual to a healthy one in  several ways~\cite{kutter},
such as through the respiratory system in the form of droplets
or aerosols  or via direct contact
(touching contaminated hand rail, hand shake, etc).
The determination of the abundance of virus  in the air~\cite{mpan}, 
their effectiveness to infect~\cite{bynumbers}, their survivability
on the surface of different types of materials~\cite{doremalen}  and
contrasting among the routes of transmissions
remain a big challenge, therefore, these factors limits our ability to
evaluate the risk~\cite{Nazhu}.
Apart from coughing and sneezing
the release of virus through respiration~\cite{richard,herfst}
and speaking is well known~\cite{asadi,stadnytskyl}.
Interestingly, it has been pointed out in~\cite{asadi}
that a large numbers of droplets carrying pathogens can
emit through human speaking and the emission intensifies
with the loudness of speech and such mechanism of
emission though independent of language spoken but
depends on some unknown physiological factors
varying among individuals.
The statistical mechanics and fluid dynamics play crucial
roles in understanding the propagation of the droplets.
Fluid dynamical tools have been applied to understand the aeorsolization 
and propagation of human droplets~\cite{dbouk,mittal}. 
The techniques of the stochastic statistical
mechanics becomes useful particularly for the
study of the motion of aerosols (droplets
with diameter  <5 $\mu$m~\cite{seta}) for which the airborne transmission
turns out to be very vital.
The aerosols undergo random Brownian or diffusive
motion in the air which can be studied within the scope of Langevin
differential equation as it contains a stochastic source term which
is normally ignored in Eulerian-Lagrangian approach~\cite{dbouk}.

In the present work we investigate 
the space-time evolution of these droplets  by taking 
into account the diffusive force through the Langevin equation. 
The diffusive force plays a crucial role particularly, 
for the motion of small droplets in the air. 
This will help enormously in planning the preventive strategies of the virus
carried by the droplets.  
The motion of the droplet ejected in the air with some initial velocity  
at some spatial point will interact with the molecules of the air.
The problem will not only be complex but unsolvable
if one considers the interaction of the droplet  with
the individual molecules of the air which are changing positions 
continuously, resulting in continuous change in the interacting force.
In such situation the air molecules can be regarded as forming a thermal bath 
characterized by temperature and density where the droplets are in motion.
The interaction
of the droplets with the bath can then be lumped into an effective force
which contains  drag and diffusive terms. 
Therefore, the interaction of
the droplets with the air can be taken into accounts through its
drag and diffusion coefficients.  
The facts stated above set an appropriate stage to study the propagation of sneezed 
and coughed droplets in the air within the scope of 
Langevin stochastic differential equation of statistical mechanics~\cite{Reif,pathria}.
It is crucial to note that the Langevin equation can be applied to solve the 
problem under study because the mass of the droplets are 
much higher than the mass of oxygen and nitrogen molecules present in the air.  
After the ejection the change in position of the
droplets with time  will be governed by the: (1) drag force exerted by the air on the droplet,
(2) diffusive force  and (3)  gravitational force acting on them. 
The inclusion of all these forces enable us to study the trajectories of droplets with
a wide range of sizes.  
The climatic conditions affect the transmission of droplets in the air (see~\cite{pica} for
details).  
The thermophysical properties of the air vary from place to place depending on the temperature
and relative humidity. These  variations have been taken into consideration 
through the temperature~\cite{ttable} and relative humidity~\cite{humidity}
dependence of the viscosity of the air. 
The viscosity of the air has been used to estimate the 
drag coefficient by employing the Stokes' formula~\cite{Landau}. 
The Einstein fluctuation-dissipation relation~\cite{pathria}
has been used to calculate the diffusion coefficient which is
directly proportional to the temperature. 
Therefore, the temperature and humidity dependence of the space-time
evolution of the droplets enter the calculation through
the drag and diffusive forces exerted by the air on the droplets.

The trajectories of the droplets will be different in still and  
flowing air [such as in a air conditioned (AC) room]. 
The propagation of the droplets in quite indoor~\cite{szhu} is very different
from a room with AC ventilation. 
The direction of air flow due to AC ventilation  plays vital role ~\cite{jlu,kang}.
Present study considers both the cases - situations with still air and wind flow.
The flow of air has been taken into account 
by using the Galilean transformation of the Langevin equation.
The velocity of the droplets  will dissipate in the air in course of time. 
It is expected that the gravitational force  is superior to both the drag and diffusive 
forces for large (massive) droplets. However, for smaller droplets drag and diffusive forces will
predominate.
Therefore, it will be interesting to study how these competitive forces 
influence the distance that the droplets traverse from the source (infected individual)
and for how long they remain suspended in the air. This
will  indicate the distance (both geometric and temporal) that  
a healthy individual should maintain from an infected one to prevent virally transmitted
diseases. 

\section{Methods - solving the Langevin equation numerically by Monte-Carlo technique}
We write down the Langevin equation  below for the motion of the droplets of mass ($M$) 
in the still air  in the presence of gravitational field~\cite{Reif}: 
\begin{equation}
\frac{dr_i}{dt}  = v_i 
\label{eq1}
\end{equation}
\begin{equation}
M \frac{dv_i}{dt} =  -\lambda v_i + \xi(t) +F^G 
\label{eq2}
\end{equation}
where $dr_i$ and $dv_i$ are the shifts of the coordinate and velocity in 
each discrete time step $dt$, $i$ stands for the Cartesian components of the 
position and velocity vectors. The 
$\lambda$ in Eq.~\ref{eq2} is the drag coefficient. 
The first term in the right hand side of Eq.~\ref{eq2} represents 
the dissipative force and the second term 
stands for the diffusive (stochastic) force where 
$\xi(t)$ is regulated by the diffusion coefficient $D$.
$\xi(t)$  is  also  called  noise due  to  its  stochastic  nature.
We study the evolution with a white noise ansatz for $\xi(t)$, {\it i.e}  
$\la{\bf \xi}(t)\ra = 0$ and
$\la {\bf \xi}(t) {\bf \xi}(t')\ra = D\delta(t-t')$. 
White noise describes a fluctuating field without memory, whose correlations have
an instantaneous decay, called  $\delta$ correlation. 
The third term in Eq.~\ref{eq2}, $F^G$ represents the gravitational
force $(=Mg$, $g=9.8$ m/s$^2$) acting on a droplet of mass $M$.

The  Galilean transformation has been used to 
take care of the flow of air (with velocity $u(x)$) into the Langevin equation. 
In the present work our aim is to
study how the dynamics of droplets are affected by the flow of the air.
Therefore, we conceive a velocity profile for the air flow  
as: $u(x)=u_0(1-\frac{x}{x_{\text{max}}})$ to serve this purpose, 
where  $x$ is the running coordinate,
$u_0$ is the peak value of $u(x)$ at $x=0$ (position of AC, say)  
and $x_{\text{max}}$ is the maximum value of $x$, which may be constrained by the 
size of an AC room. However,  more complex velocity profile can also
be contemplated.  We have taken the flow velocity along horizontal direction 
with vanishing components along upward and downward  directions. 
It is obvious that any non-zero upward (downward) component will enhance
(reduce) the time of suspension of droplets in the air.

We solve the Langevin equations, Eqs.~\ref{eq1}, \ref{eq2} 
simultaneously by using Monte-Carlo techniques~\cite{mc1,mc2} with the inputs discussed below. 
For the initial spatial coordinate we use, $x=y=0$ and $z=H_0$, where $H_0$ is the height
(taken as 1.7 meter) at which the droplet is released (nose/mouth),
that is the initial spatial coordinate of the droplet is $(x,y,z)=(0,0,1.7{\text {meter}})$.  
We distribute the initial velocity uniformly in the 
$x-y$ plane, where $v_z=0$. The gravitational force act on the downward $z$ direction.
We vary  the  radius ($R$) of the droplets from 2.5$\mu$m to 100$\mu$m~\cite{RR} and
the ejection velocity ($V_0$) from 5 to 21 m/s~\cite{VE}. 
The mass  of the droplet has been estimated from the radius ($R$)
by using the relation 
$M=4\pi R^3d/3$, where $d(=997$ kg/m$^3$) is the density of droplet. 
The value of the drag coefficients, $\lambda$ is estimated by using 
the relation, $\lambda=6\pi\eta R$, obtained from Stokes formula. 
The value of the diffusion coefficient is obtained by using the Einstein relation
\cite{pathria}, $D=K_BT\lambda$,
where $K_B=1.38\times 10^{-23} J/^{\circ}K$, is the Boltzmann constant and $T$ is 
the temperature.
We consider $L(t)=\sqrt{x(t)^2+y(t)^2}$ as the horizontal distance traveled by the droplet from the 
point of ejection and the maximum value of $L(=L_{\text{max}})$ dictates the stationary
distance that to be maintained between infected and healthy persons to avoid the virus. 

It may be mentioned that if we set the values of  drag and diffusion coefficients 
to zero then our numerical results are in excellent agreement with the results obtained
by assuming free fall of the droplets with large size (mass).  

\section{Results}
\begin{figure}[t]
        \centering
        \includegraphics[width=12cm, height=8cm]{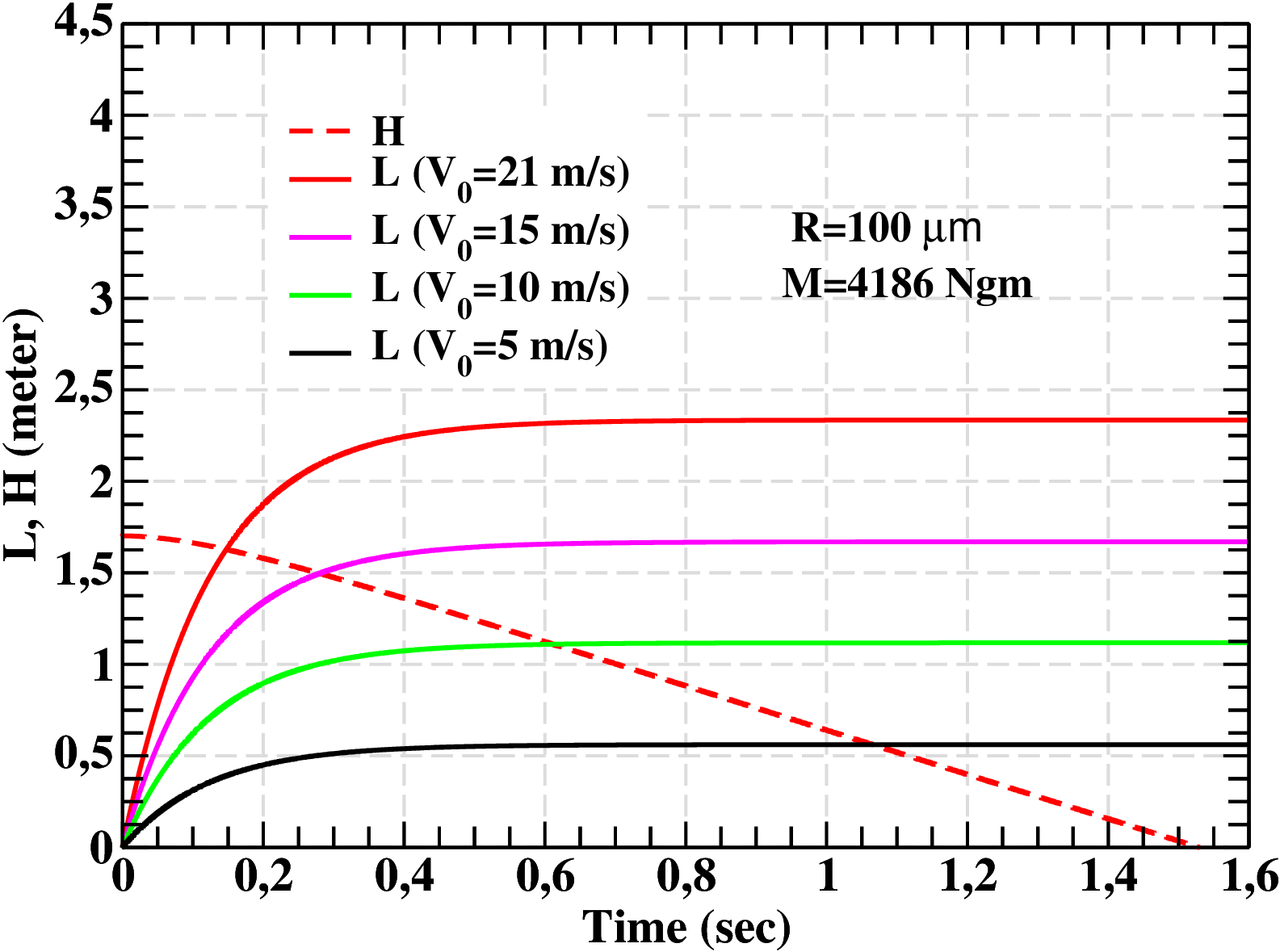}
        \caption{The  horizontal distance, $L(t)$ traveled by the ejecta of mass
4186 Ngm (nano-gm) from the 
source of infection as a function of time for different ejection velocities
have  been shown here. The droplets are
ejected at a height 1.7 meter from the ground.  
The change of  height, $H(t)$  with time of the droplet for 
initial velocity 21 m/s has also been depicted. 
}
        \label{Fig1}
        \end{figure}

\begin{figure}[t]
        \centering
        \includegraphics[width=12cm, height=8cm]{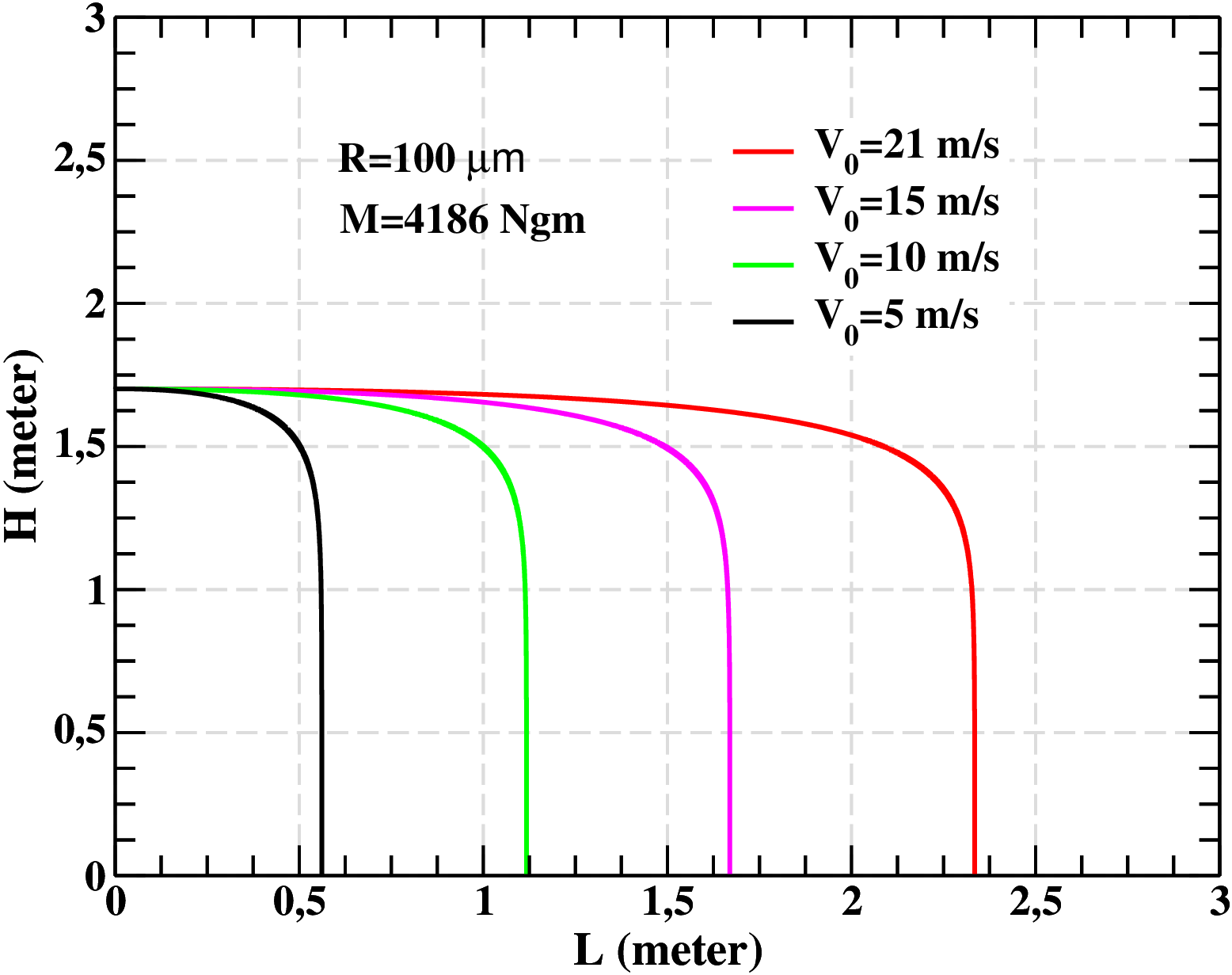}
        \caption{Variation of $H$ with $L$ for a droplet of mass 4186 Ngm and radius 100 $\mu$m for
different ejection velocities have been displayed.
                 } 
        \label{Fig2}
        \end{figure}

Among other factors,
 the contamination depends on the mass and initial velocity of the droplets. However, 
the droplets ejected through coughing and sneezing will have  different 
sizes (and  hence masses) and  initial  velocities. Therefore, we provide results 
for a range of droplet sizes and initial velocities.  The contagion by the droplets
will  also depend
on the air flow, temperature and humidity of the air where the droplets are discharged.
Sensitivities of the results on these factors have also been investigated and discussed
below.
The results presented in Figs.~\ref{Fig1} and ~\ref{Fig2} have been obtained
in still air at temperature, $T=30^\circ$C with inputs discussed above. 
In Fig.~\ref{Fig1} the time variation of horizontal distance ($L$) travel by droplets 
for various ejection velocities have been displayed.  
The (horizontal) distance, $L$ traveled by the droplets from the source depends 
strongly on the initial velocity and mass. While a droplet of mass 4186 Ngm 
with small ejection velocity, $V_0=5$ m/s travels a distance, $L\approx 0.55$ meter, 
a droplet with larger $V_0=21$ m/s travels 2.35 meter 
approximately. This droplet takes about 1.5 sec before it settles 
on the ground under the action of
gravity. Other droplets with intermediate values of 
$V_0=$ 15 m/s and 10 m/s travel horizontal distances approximately
1.7 meter and 1.1 meter respectively.
It may be mentioned here that a droplet of radius 200 $\mu$m takes about 
0.73 sec to fall on the ground 
under the action of gravity which may be compared with the value for free fall time 
($t=\sqrt{(2H)/g}=0.59 sec$) from a height 1.7 meter (please also see ~\cite{FF,LAC}). 
This indicates that free fall under gravity will be a reasonable approximation 
for droplets having radii larger than 200 $\mu$m. The red dashed line in Fig.~\ref{Fig1} shows the variation of height, 
$H(t)$  with time when it is released at an initial height, $H_0=1.7$ meter with $V_0=21$ m/s.  
The time variation of $H(t)$ for large droplets (mass 4186 Ngm or more)
with other values of $V_0$ are not shown because of its weak $V_0$ dependence.
The results discussed above can be viewed in a different way as follows. 
Fig.~\ref{Fig2} shows the change of height ($H(t$) of the droplets
with horizontal distance ($L(t)$) for different initial 
velocities ($V_0$). A droplet of mass 4186 Ngm with $V_0=21$ m/s (5 m/s) travel a 
horizontal distance, $L\approx 2.35$ meter (0.55 meter).   
The same droplet with intermediate $V_0$ values, 15 m/s (10 m/s) travels 
approximately 1.7 meter (1.1 meter). These results indicate that big (massive)
droplets falls on the ground within a short time due to gravitational  force
but they travel larger distance due to larger momentum as the drag force for such
droplets is weaker than gravitational force.  These results are consistent with 
the results displayed in Fig.~\ref{Fig1}. 
 \begin{figure}[t]
        \centering
        \includegraphics[width=12cm, height=8 cm]{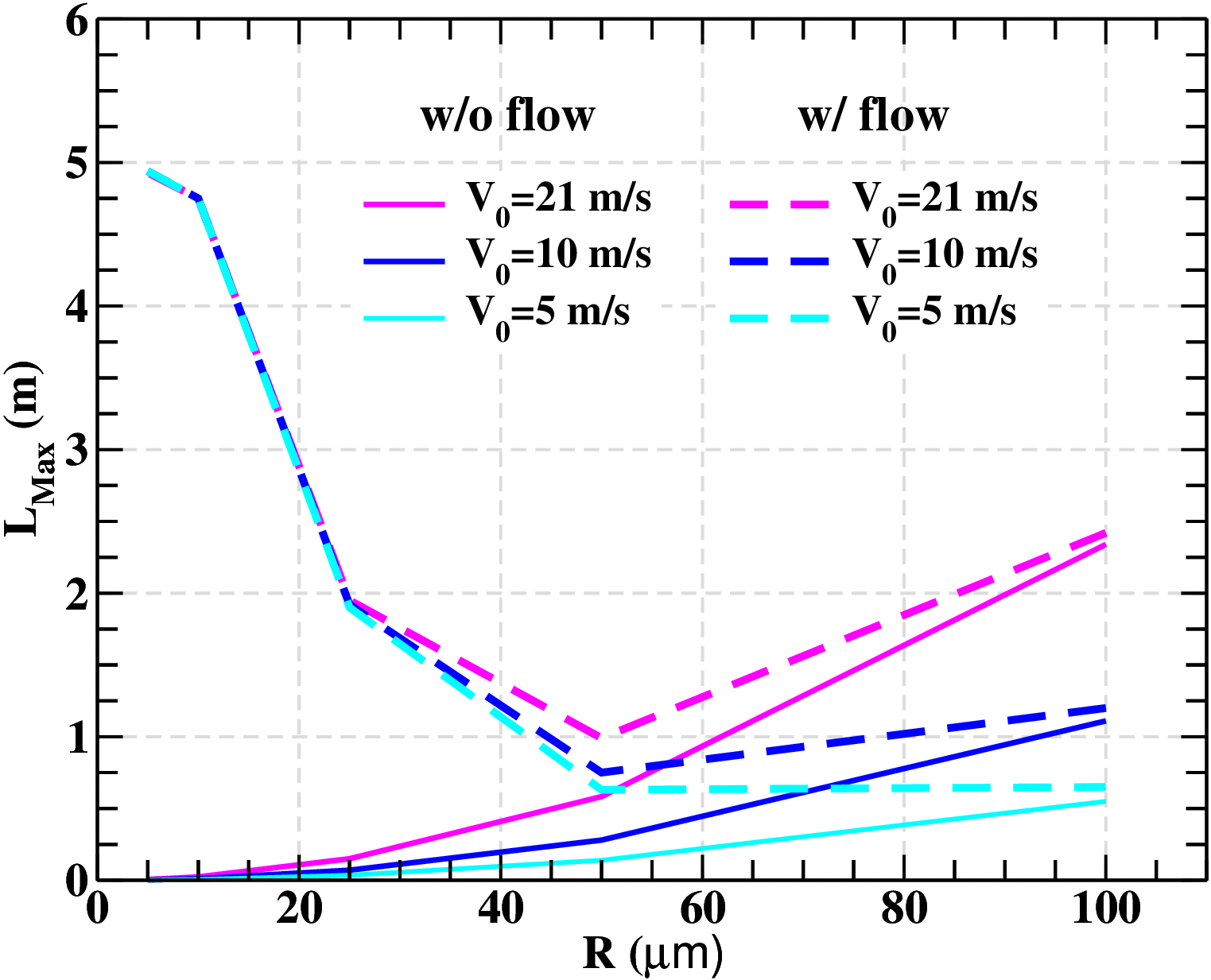}
        \caption{The variation of the maximum horizontal distance ($L_{max}$)
          traveled by droplets as a function of radius for different ejection velocity.}
        \label{Fig3}
        \end{figure}
From the preventive strategic point of view the question to be asked
is - what is the maximum horizontal distance ($L_{\text{max}}$) that a healthy individual should
maintain from an infected one? The answer will depend on several factors discussed
above, {\it e.g.} ejection velocity, 
mass of the droplets, temperature, humidity, flow velocity 
of the air, etc. We evaluate $L_{\text{max}}$ both in still and flowing air conditions 
and display its variation with the radius ($R$) of
the droplets for ejection velocities, $V_0=5, 10, 21$ m/s 
in Fig.~\ref{Fig3} at a temperature, $T=30^\circ$C. 
It is appropriate to mention here that the mean value of $V_0$ is about 10 m/s and 
the value 21 m/s is 
close to the highest possible value of $V_0$ for  droplets originating from coughing
~\cite{VE,zhu}. 
The value of $u_0$ appearing in the velocity profile of the wind 
mentioned above has been 
taken as, $u_0=0.1$ m/s and $x_{\text{max}}=5$ meter.

We observe that the maximum horizontal distance traveled
by the droplets in still air increases with its size or mass (Fig.~\ref{Fig3}). 
Droplets with larger $V_0$ gives  larger  value of $L_{\text{max}}$ for given $R$.
It is crucial to note that $L_{\text{max}}$ for large (massive) droplets does not change much
with moderate air flow.  The action of gravity on large droplets dominates over 
the drag and diffusive forces
and hence they expeditiously settle gravitationally. 
In still air a droplet of radius  $100\,\mu$m travels 2.35 meter, 1.1 meter, 0.55 meter for 
$V_0=21, 10$ and 5 m/s  respectively. 
The drag and diffusive forces does not allow small droplets to travel long distance in still air.

However, in flowing air condition the scenario is very different. 
Gravitational force imparts a   
downward terminal velocity ($v_t$) to the droplet which is given by: 
$v_t=2R^2(d-\rho)g/(9\eta)$,
where $\rho$ is the density of air.  In an environment of flowing air with flow velocity $u(x)$,
the resultant of $v_t$ and $u$ will dictate how long a droplet will move
before gravitationally settled.  
If $v_t$ of a droplet  is large compared to $u_0$ (peak value of the
flow velocity) then it will quickly settle under
the action of gravity. 
The value of $v_t$ for a 100$\mu$m droplet is 1.2 m/s which is 12 times larger than the 
peak value of the flow velocity,  $u_0$ ($=0.1$ m/s), therefore, such droplets 
will strike the ground fast without much effects of flow.  
However, the smaller droplets are strongly affected by the flow.
The value of $v_t$ for a $5\mu$m droplet is $0.3\times 10^{-2}$ m/s which
is more than an order of magnitude lower than  $u_0(=0.1$ m/s). 
Such small droplets are influenced by drag, diffusion and flow and
have more time to travel large distances~\cite{Bourouiba}. A droplet of radius 5 $\mu$m will
traverse a distance 4.95 meter. 
It is crucial to note that for
small droplets, $L_{\text{max}}$ is insensitive to ejection velocity.
Droplet with intermediate size experience some sort of cancellation between the actions
of gravitational and drag force and therefore move smaller distance 
if flow velocity is low compare to the corresponding values of their 
$v_t$~(Fig.\ref{Fig3}).

We have also considered $u_0=0.25$ m/s to understand the effect of air flow. We found
that the increase in  flow velocity from 0.1 m/s to 0.25 m/s changes the
$L_{\text{max}}$ by $1\%$, $88\%$ and $8.2\%$ for droplet of radii 5 $\mu$m,
50$\mu$m and 100$\mu$m respectively. It is clear that the effect of increase of $u_0$ 
on a $5\mu$m droplet is small because its $v_t<<0.1$ m/s 
and hence any further increase in $u_0$ has negligible influence.
A  5$\mu$m droplet will travel a distance of 4.95 meter (5 meter) 
from the point of ejection if the peak flow velocity is 0.1 m/s (0.25 m/s). 
Similarly for a 100$\mu$m droplet the gravitational effect still dominates 
because their $v_t$ is more than $0.25$ m/s, 
resulting in only about 8.2$\%$ increase in $L_{\text{max}}$.
However, a $50\mu$m droplet has $v_t=0.3$ m/s which is comparable to $u_0=0.25$ m/s
and hence the change in $L_{\text{max}}$ for such droplet is substantial ($88\%$).   
Therefore, it is important to note that the distance traveled by
a droplet will depend on the interplay  between the magnitudes of 
downward terminal velocity and the flow velocity.
Therefore, as preventive strategies a healthy person should maintain
different distances in still and flowing air environments. 

We note that for smaller droplets $L_{\text{max}}\approx x_{\text{max}}$,
suggesting that the dynamics of these droplets is almost entirely determined 
by the air flow. At a distance, $L=x_{\text{max}}$
the velocity profile of the air turns into zero and the drag of the air becomes 
dominant which does not allow the smaller droplets to travel anymore.
 \begin{figure}[t]
        \centering
        \includegraphics[width=12cm, height=8 cm]{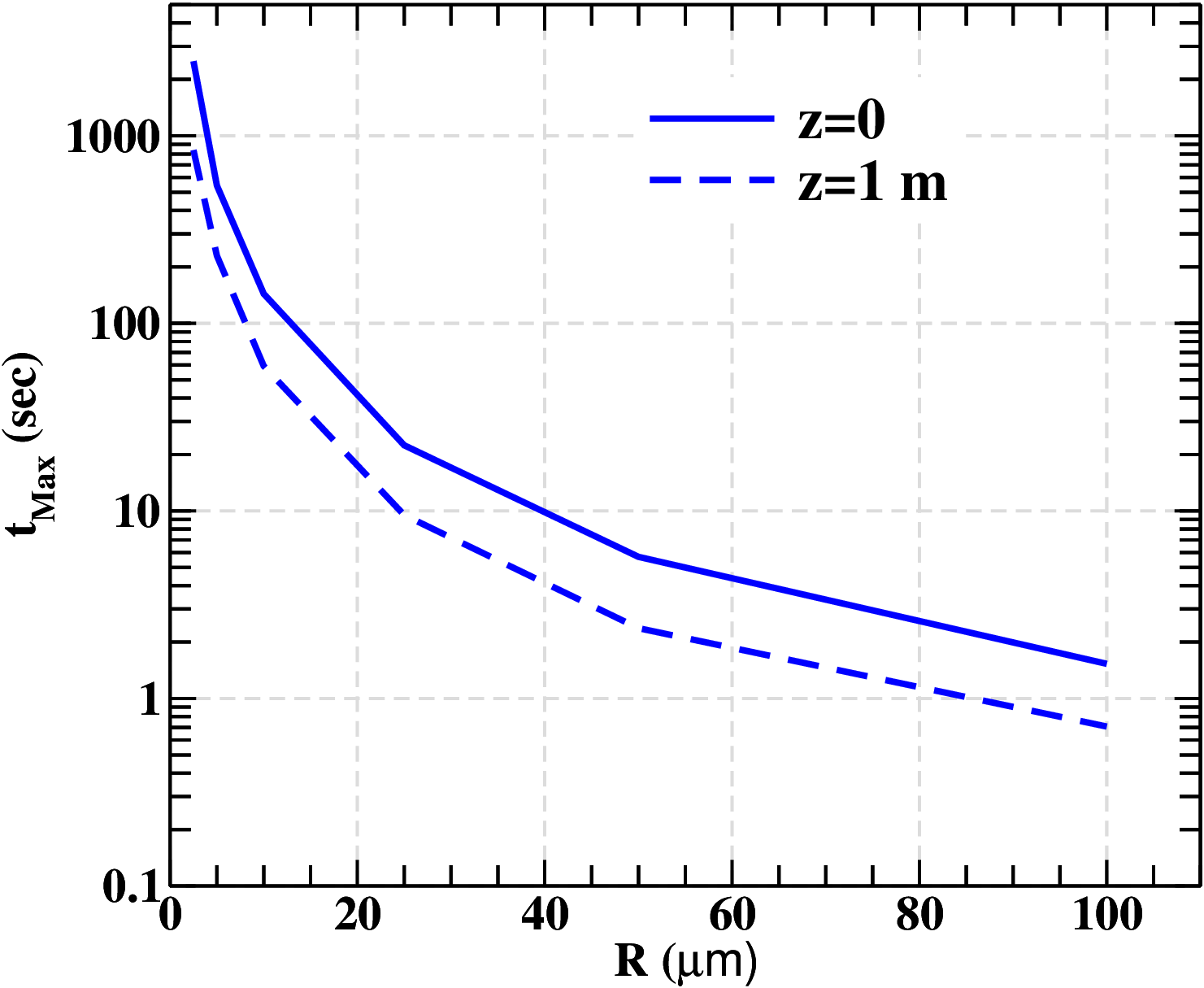}
        \caption{The variation of the maximum times the droplets take to
reach a height of 1 meter from the ground (dashed line) and to hit the ground (solid line).
                }
        \label{Fig4}
        \end{figure}

How long a droplet takes to gravitationally settle on the ground or in other words
how long it remains suspended in the air after it is ejected through sneezing
or coughing? 
In Fig.~\ref{Fig4} the maximum time ($t_{\text{max}}$) of suspension of the droplet in the air 
is plotted as a function of $R$ for $V_0=21$ m/s for $T=30^{\circ}$C.
We find that the dependence of $t_{\text{max}}$ on $V_0$ is mild. 
The results clearly indicate that 
$t_{\text{max}}$ decreases with increase in $R$ {\it i.e.} the smaller droplets remain suspended 
in the air for a longer time.
We find that a droplet of size 100 $\mu$m float in the air only for 1.5 sec approximately. 
For large (massive) droplets the gravitational force dominates over drag and diffusion
and consequently they settle on the ground quickly. However, a  droplet of smaller
size (hence lighter too) of radius 2.5 $\mu$m survives in the air for  about 41 minutes 
(Fig.\ref{Fig4}) for $u_0=0.1$ m/s  because for such lighter droplets effect of gravity 
is small.
This result may be used as a guideline to determine the temporal distance that a healthy individual
should maintain from an infected one.  
A healthy individual should not only be careful 
in maintaining the geometric distance from an infected individual
but also deter the suspended lighter droplets by suitably covering nose, mouth, etc
by using mask~\cite{dbouk,Howard,pnas,verma} and other possible accessories.
Moreover, the majority of droplets ejected from exhalation process has radius around 
10$\mu$m~(\cite{VE} and references therein) for which the use of mask is necessary.
Fig.~\ref{Fig4} also
display the maximum time taken by droplets of various sizes to fall at height of 1 meter 
from its released position at a height of 1.7 meter from the ground. 

\section{Discussions}
The maximum time of suspension in the air and the 
maximum horizontal distance traveled  by the droplets ejected 
by infected individuals through coughing and sneezing 
have been estimated both for still and flowing air conditions 
by solving the Langevin equation. 
All the possible forces (drag, diffusive and gravitational)
which influence  the dynamics of the droplets in the air
have been taken into accounts under
varying conditions of temperature, humidity and air flow. The sizes and
the initial ejection velocities used  in the calculations have been taken from
measured values available in the literature~\cite{RR,VE}. 
With all these inputs the Langevin equation has been solved 
rigorously to find that 
the small droplets travel larger distance and remain
suspended in the air for long time under the influence
of air flow making the use of mask  mandatory to prevent the virus. 
Therefore, the maintenance of only stationary separation between healthy
and infected individual is not substantiated.  Calculations based on fluid dynamics  
~\cite{Bourouiba} show that small droplets has the ability to carry the pathogens to longer 
distance which corroborates the fact the maintenance of only six feet social distancing is
not sufficient to evade  the virus.

We have studied the impact of diffusion, represented 
by the term $\xi(t)$ appearing in the right hand side of
equation Eq.~\ref{eq2} on droplets of different sizes. 
It is found that droplets with radii $\leq 5 \mu$m are affected considerably. 
For example, the time of suspension of a droplet of radius 2.5 $\mu$m in the air 
is changed by about $25\%$ as a result of diffusion.  
It is found that  smaller droplets follow zig-zag paths with slight 
variance around its  trajectory and stay longer in the air due to diffusion. However, the 
impact of diffusion on larger droplets is insignificant.

We notice that such a result goes along with the very recent finding~\cite{Liu}
that in two Wuhan hospitals the micrometer and sub-micrometer droplets
of Sars-COV-2 were found at a distance of about 3 meters 
from the infected patient's bed around to the room corners, where indeed the
air flow is damped and/or twirls into local vortices.

It may be mentioned here that smaller droplets may originate from the fragmentation ~\cite{xuwang}
or evaporation~\cite{FanLiu} of the larger droplets and remain suspended in the air for longer
time causing potential health problems. 
Again in such cases preservation of static separation is not justifiable. 
Isolated virus may be created from the
process of evaporation. The  survivability of these virus  in air for more than a hour 
has been reported~\cite{swxong}. Such virus will remain suspended in the air for long
time due to the dominant actions of  drag and diffusive forces as well as air flow as the
gravitational influence on them is weak.  
However, it is important to mention here that in a very interesting recent work ~\cite{jama} it has been
pointed out that by the processes of  sneezing and coughing not only droplets are
produced but also  multi-phase turbulent gas cloud which can carry 
cluster of droplets of all possible sizes. In such a scenario these contained droplets
can avoid the evaporation and hence can live longer than isolated droplets.

We have performed a thorough study considering also possible effects coming from the
higher order correction, associated to large Reynolds number,
with respect to the  Stokes' approximation but we have found only negligible changes for small 
droplets ($R\leq 10 \mu$m) that can be discarded at the level of accuracy relevant in this context
(for large droplets action of gravity dominates over viscous force).  
We have also studied the impact of the temperature on the space-time evolution of the trajectories
exploring a wide range from 0 to 40 $^o C$, the results shows a limited impact that is of about 
$10\%$.

\section*{Data Availability Statement}
The data that supports the findings of this study are available within the article.

\section*{Acknowledgement}  
SKD would like to acknowledge IIT Goa for internal funding (No. 2020/IP/SKD/005)
and Professor Barada Kanta Mishra for useful discussions.


\end{document}